# Unveiling the anisotropy of linear and nonlinear charge-spin conversion in Weyl semimetal TaIrTe$_4$

Tao Tang[†,§], Mengzhou Li[†,§], Bin Lao[†,§], Xuan Zheng[†,§], Wei Zhou[ω], Xiaofeng Xu[ε], Jie Pang[‡], Youguo Shi[‡], Run-Wei Li[*,†,§,ξ] and Zhiming Wang[*,†,§,ξ]

[†]CAS Key Laboratory of Magnetic Materials and Devices, Ningbo Institute of Materials Technology and Engineering, Chinese Academy of Sciences, Ningbo 315201, China

[§]Zhejiang Province Key Laboratory of Magnetic Materials and Application Technology, Ningbo Institute of Materials Technology and Engineering, Chinese Academy of Sciences, Ningbo 315201, China

[ω]School of Electronic and Information Engineering, Changshu Institute of Technology, Changshu 215500, China

[ε]School of Physics, Zhejiang University of Technology, Hangzhou 310023, China

[‡]Institute of Physics, Chinese Academy of Sciences, Beijing 100190, China

[ξ]Center of Materials Science and Optoelectronics Engineering, University of Chinese Academy of Sciences, Beijing 100049, China

**ABSTRACT**

In Weyl semimetals, the nonlinear planar Hall effect (NPHE) and spin-orbit torque (SOT) are prominent manifestations of nonlinear and linear charge-spin conversion, respectively. However, simultaneous investigations of these phenomena within a single material system are scarce, limiting our understanding of their intrinsic connection and underlying mechanisms. Here, we report the first simultaneous observation of NPHE and SOT in a TaIrTe$_4$/Py heterostructure. By employing harmonic Hall measurements and developing a magnetic field-dependent method, we successfully separated the contributions from NPHE, field-like SOT, and damping-like SOT, enabling accurate characterization of both linear and nonlinear charge–spin conversion properties. Our experiments revealed significant anisotropy along the [100] and [010] crystallographic directions of TaIrTe$_4$, with stronger nonlinear responses and field-like SOT along the [100] direction, and larger damping-like SOT along the [010] direction. The distinct directional dependence of these phenomena provides new insights into the interplay between surface and bulk contributions to charge-spin conversion in Weyl semimetals. These findings enhance our understanding of anisotropic charge-spin conversion mechanisms in Weyl semimetals, which may inform future research and development of spintronic devices based on topological materials.

**Introduction**

Spintronic devices hold great promise for revolutionizing the microelectronics industry by integrating storage, sensing and logic functions, offering advantages such as low power consumption, high computing speed and non-volatility.[1-4] A central challenge in spintronics is achieving effective spin manipulation to enable efficient and controllable charge-spin conversion. Compared to methods like spin injection,[5] spin pumping,[6] and other techniques that generate spin currents through ferromagnetic metals, utilizing spin-orbit coupling (SOC) can achieve more efficient charge-spin conversion with a diverse range of underlying mechanisms. The spin Hall effect (SHE)[7] can be observed in materials with strong SOC (e.g., heavy metals), resulting in spin currents flowing along the transverse direction of the charge current. Moreover, various charge-spin conversion phenomena, such as Rashba effect[8, 9] and the surface states of topological materials,[10-15] have been observed in certain interface and surface systems with SOC. Gaining a clear understanding of SOC and its corresponding bulk and interface/surface effects is crucial for investigating the physical mechanisms underlying charge-spin conversion and developing strategies for enhancing the charge-spin conversion efficiency.

Charge-spin conversion can be categorized into linear and nonlinear responses based on how the spin current depends on the charge current.[16, 17] Linear charge-spin conversion, exemplified by phenomena such as spin-orbit torque (SOT),[18-20] relies on a linear response relationship between the spin accumulation and the applied electric field and is fundamental for manipulating spins in spintronic devices.[21] Nonlinear charge-spin conversion, on the other hand, arises in materials with strong SOC and broken inversion symmetry, leading to effects like the nonlinear planar Hall effect (NPHE).[10, 22-24] Although both linear and nonlinear mechanisms are crucial for understanding spin current generation and enhancing charge-spin conversion efficiency, studies that simultaneously investigate both within a single materials system are scarce. This lack of combined studies limits a comprehensive understanding of the underlying microscopic mechanisms and their interrelation.

Weyl semimetal (WSM), such as TaIrTe$_4$, stands out as a promising candidate for studying charge-spin conversion mechanisms. It possesses only two pairs of well-separated Weyl points,[25, 26] the minimum number allowed under time reversal symmetry conditions, making it particularly suitable for theoretical calculations and experimental analysis.[27, 28] The surface state structure of the Fermi arc of TaIrTe$_4$ may exhibit spin polarized states[29] similar to spin-momentum locking, which can generate a second-order charge-spin conversion.[30, 31] Additionally, TaIrTe$_4$ exhibits a large Berry curvature at the Weyl point.[26, 32] These intriguing bulk and surface electronic states make TaIrTe$_4$ an ideal material platform for studying the intricate mechanism of charge spin conversion.

In this work, we aim to bridge the gap in understanding the intrinsic connection between linear and nonlinear charge–spin conversion mechanisms by simultaneously investigating both phenomena within a single material system. We fabricate TaIrTe$_4$/Py heterostructures as our experimental platform to investigate both phenomena. By developing a magnetic field-dependent harmonic Hall measurement technique, we successfully separate the contributions from the NPHE, field-like SOT, and damping-like SOT, allowing for a comprehensive investigation of the charge–spin conversion properties along different crystallographic directions ([100] and [010]) in TaIrTe$_4$. Our systematic approach provides new insights into the role of spin-momentum locking and anisotropic topological surface states in both linear and nonlinear charge-spin conversion processes.

**Results and Discussion**

**A. Characterizations of TaIrTe$_4$ single crystals**

Figure 1(a) illustrates the crystal structure of TaIrTe$_4$ from the front (left) and top (right) views. Alternating Ta–Ir connections stretch along the crystalline a-axis ([100]) as zigzag atomic chains. To investigate the properties of TaIrTe$_4$, flakes were mechanically exfoliated[33] from bulk single crystals and transferred onto a silicon substrate with 500 nm SiO$_2$, as shown in Figure 1(b). The representative flake thickness, as measured by atomic force microscopy (AFM), is 117 nm [Figure 1(c)], which is suitable for device fabrication and transport measurements. The crystalline quality and orientation of the TaIrTe$_4$ single crystals were characterized by x-ray diffraction (XRD). Figure 1(d) displays the XRD pattern of the TaIrTe$_4$ single crystal, indicating that the cleavage surface of the crystal is parallel to the (001) plane and confirming the absence of impurity phases. To further investigate the anisotropic properties of TaIrTe$_4$, angle-dependent polarized Raman spectroscopy was employed. Figure 1(e) presents the typical angle-dependent spectra of a TaIrTe$_4$ flake, which were measured in a parallel-polarized configuration. The angular dependence of the intensity of the most prominent Raman peak, which is attributed to the A$_g$ mode of TaIrTe$_4$ and is located at a frequency of 86 cm$^{-1}$, is presented in Figure 1(f). This peak exhibits a clear two-fold symmetry, with maximum intensities observed when the polarization of the incident light is parallel to the [100] direction of the TaIrTe$_4$ flake.[34]

**B. Second harmonic Hall measurements**

Having established the high quality and anisotropic nature of our TaIrTe$_4$ crystals, we proceed to investigate the spin-orbit torque (SOT) effects in the TaIrTe$_4$/Py bilayer system using harmonic Hall measurements.[35-37] In such bilayer structures, SOT can arise from two main mechanisms: the Rashba-Edelstein effect (REE) at the interface[38,39] between the non-magnetic (NM) and ferromagnetic (FM) layers, and the spin Hall effect (SHE) in the bulk[40-42] of the NM layer. These effects generate field-like (FL)[43-45] and damping-like (DL)[46-48] torques on the FM layer magnetization.

Experimentally, harmonic Hall measurements are widely adopted in the community for evaluating SOT efficiency.[21] The samples used for harmonic Hall measurements are typically micron-sized Hall bars fabricated from the NM/FM bilayer as shown in Figure 2(a). An external magnetic field and an AC current (a few kHz) are applied to the Hall bar, and the first and second order harmonic responses of the Hall voltage are read out using a standard lock-in amplifier. By varying the external magnetic field (strength and direction) and analyzing the Hall voltage response, it is possible to extract the longitudinal and transverse effective fields (caused by the DL and FL torques, respectively).[35-37]

Interestingly, TaIrTe4 being a Weyl semimetal (WSM), exhibits an additional phenomenon known as nonlinear charge-spin conversion,[30, 31, 49] which is closely related to the nonlinear planar Hall effect (NPHE). As shown in Figure 2(c), when an external magnetic field H is applied parallel to the electric field E, a transverse nonlinear spin current is generated. This spin current can be partially converted back into a charge current $J_c(E^2)$, which is detectable through harmonic Hall voltage measurements. The amplitude of $J_c(E^2)$ varies with the angle φ between E and H, following a cosφ dependence.[10, 16, 50-52] This angular variation manifests as the NPHE, which bears a striking similarity to the angular dependence of the damping-like torque ($H_{DL}$) in FM layers with in-plane magnetic anisotropy, despite arising from fundamentally different mechanisms.

More critically, in the FM/WSM bilayer system, the $R_{xy}^{2\omega}$ signal contains both the magnetization precession response from the FM layer and the nonlinear charge-spin conversion from the WSM layer. To separate these contributions and quantitatively characterize both SOT and NPHE effects, we employed the following formulas:

$$R_{xy}^{2\omega} = -R_{FL+Oe}\cos\varphi\cos2\varphi - \frac{1}{2}R_{DL+NPHE}\cos\varphi \quad (1)$$

$$R_{FL+Oe} = \frac{R_P(H_{Oe} - H_{FL})}{H} \quad (2)$$

$$R_{DL+NPHE} = \frac{R_A H_{DL}}{H_K + H} + R_{xy}^{\nabla T} + 2\chi_{NPHE}H \quad (3)$$

$R_{FL+Oe}$ and $R_{DL+NPHE}$ represent the resistances extracted from the different angle dependencies, with $R_P$ and $R_A$ being the planar Hall resistance and anomalous Hall resistance, respectively. $H_K$ is the effective out-of-plane anisotropy field and $H_{Oe}$ is the Oersted field estimated using Ampere's law. $R_{xy}^{\nabla T}$ represents the second harmonic transverse resistance from thermal effects (the anomalous Nernst effect and spin Seebeck effect),[53-55] while the last term in Eq.(3) describes the NPHE contribution, showing linear dependence on H.

Our analysis reveals distinct field-dependent behaviors in the angular dependence of $R_{xy}^{2\omega}$. Figure 2(d) displays the typical φ dependent second harmonic Hall voltage

$R_{xy}^{2\omega}$ measured at low fields, including the fitted curves of the $R_{FL+Oe}$ and $R_{DL+NPHE}$ components. Due to the presence of $H_K$ (a few kOe) in the first term of Eq. (3), the value of $1/H$ will be larger than $1/(H_K - H)$, and considering the comparable coefficients, the $R_{FL+Oe}$ dominates the signal in this regime. As shown in Figure 2(e), however, the signal becomes dominated by the $R_{DL+NPHE}$ at large fields. This transition occurs because, although the DL-SOT contribution decreases at large fields, the overall magnitude of $R_{DL+NPHE}$ gradually increases due to the linear NPHE term in Eq. (3). This observation presents an apparent contradiction: the increase of signal with field is inconsistent with the expected negative correlation between $R_{DL}$ with H. As shown in Figure 2(f), this unexpected behavior strongly suggests that the NPHE in the WSM layer significantly contributes to the second harmonic Hall signal, particularly at high fields. Therefore, to accurately characterize the SOT efficiencies in the FM/WSM bilayer using the second harmonic Hall technique, it is crucial to eliminate the NPHE signal from the measurements.

## C. Nonlinear planar Hall components

Having demonstrated the significant NPHE contribution at high fields, we now focus on characterizing this nonlinear response in detail. As shown in Figure 3(a) and (b), $R_{xy}^{2\omega}$ exhibits a clear cosine dependence on $\varphi$ at fixed current and magnetic field values, where the SOT contribution becomes negligible compared to the NPHE signal. The amplitude of the second harmonic Hall resistance $\Delta R_{xy}^{2\omega}$ gradually increases under a fixed I of 2 mA with various B from 1 T to 6 T and under a fixed H of 5 T with various I from 1.4 mA-2.2 mA. To intuitively reflect the nonlinear transport properties of TaIrTe$_4$, we extracted the $\Delta R_{NPHE}$ from the data using the relation $R_{xy}^{2\omega} = \Delta R_{NPHE} cos\varphi$ and investigate the relationship between $\Delta R_{NPHE}$ and the input current I and the in-plane magnetic field H. In Figure 3 (c) and (d), $\Delta R_{xy}^{2\omega}$ is linearly related to both I and H, indicating that the nonlinear planar Hall effect can be observed when the current flows along the [100] crystal direction of TaIrTe$_4$, which reflects the special topological surface states.[26, 29] To further explore the anisotropic nature the NPHE in TaIrTe$_4$, a current was then applied along the TaIrTe$_4$ [010] direction (Supporting Information Note 1), and the $\Delta R_{NPHE}$ data were processed in a similar way. Remarkably, $\Delta R_{NPHE}$ is also linearly related to both I and H when the current is applied along the [010] direction, confirming that the TaIrTe$_4$ [010] direction is also capable of producing a nonlinear planar Hall effect. It is crucial to acknowledge that the NPHE signal for charge current flowing along the [100] direction is, in fact, a response to the spin-momentum locking nature along the [010] direction, highlighting the complex interplay between the crystal symmetry and the nonlinear transport properties in TaIrTe$_4$.

**D. Spin orbit torque components**

After characterizing and accounting for the NPHE contribution at high fields, we now focus on analyzing the SOT components in the low-field regime where they become dominant. This analysis is crucial for understanding the linear charge-spin conversion mechanisms in TaIrTe$_4$. As shown in Figure 4(a), the first harmonic Hall voltage $V_{xy}^{1\omega}$ of the TaIrTe$_4$/Py bilayer follows the sin2φ dependence with I = 2 mA and H = 100 Oe, suggesting that the magnetization of the Py layer is consistently aligned with the external magnetic field for H > 100 Oe. In Figure 4(b), the typical $\varphi$ dependent second harmonic Hall voltage $V_{xy}^{2\omega}$ for a change in magnetic field from 100 Oe to 500 Oe is displayed at a fixed I = 2 mA. The data maintains the angular dependence of $cos\varphi cos2\varphi$, consistent with our previous analysis. According to Eq. (1), in this low field region, the signal is primarily contributed by the FL-SOT and the Oersted field.

As shown in Figure 4(c), increasing the magnetic field from 500 Oe to 20000 Oe causes the angular dependence to evolve from $cos\varphi cos2\varphi$ to $cos2\varphi$, indicating a transition in the dominant contribution to the signal. Following Eq. (1), we can separate the $V_{FL+Oe}$ and $V_{DL+NPHE}$ terms through fitting analysis. Figure 4(d) shows the $V_{FL+Oe}$ for I varying from 1.4 to 2.2 mA as a function of 1/H. After removing the

Oersted field contribution, each $V_{FL}$ follows a linear relationship with the current density j of the TaIrTe$_4$ layer. Using Eq. (2), we extract the effective field $H_{FL}$ from these relationships as shown in Figure 4(f). Similarly, we obtain the effective field $H_{DL}$ using Eq. (3) by carefully removing the linear NPHE term in the appropriate magnetic field range (250 Oe to 5000 Oe). Using the formula $\theta_{SH} = \frac{2e\mu_0 M_s t H_{FL(DL)}}{\hbar J}$,[56] we estimated the FL-SOT efficiency $\theta_{FL}$ to be 0.57 and DL-SOT efficiency $\theta_{DL}$ to be 0.93, where $M_s$ and $t$ are the saturation magnetization and the thickness of the Py layer, respectively. Similar measurements were also performed with current flowing along the [010] direction (Supporting Information Note 2), enabling a comprehensive investigation of SOT anisotropy to be discussed in the following section.

**E. Anisotropic nonlinear planar Hall effect and spin-orbit torques**

Having separated the NPHE and SOT contributions, we now investigate their crystallographic direction dependence in TaIrTe$_4$. Our measurements demonstrate pronounced anisotropy in the nonlinear transport properties, as shown in Figures 5(a). To quantitatively characterize the anisotropic behavior, we analyze the nonlinear Hall coefficient $\chi = \rho_{xy}^{2\omega}/(E_x H_y)$ along different crystallographic directions. Along the [100] direction $\chi$ reaches is 0.74 mΩ·V$^{-1}$·µm$^2$·T$^{-1}$, while along the [010] direction, it decreases to 0.15 mΩ·V$^{-1}$·µm$^2$·T$^{-1}$. The five-fold difference in $\chi$ demonstrates the significant anisotropic nature of nonlinear charge-spin conversion in TaIrTe$_4$. The SOT measurements, presented in Figures 5(b) and (c), reveal similar striking directional dependence: the FL-SOT efficiency shows stronger response along the [100] direction, while the DL-SOT efficiency exhibits enhanced behavior along the [010] direction, suggesting different underlying mechanisms for these two types of torques.

Notably, we observe a positive correlation between NPHE and FL-SOT responses along the [100] direction, suggesting their shared microscopic origin. This correlation is particularly significant as it indicates that similar mechanisms might govern both the nonlinear charge-spin conversion and field-like torques. In contrast, the DL-SOT shows stronger response along [010], pointing to a different underlying mechanism. The observed anisotropic behavior can be understood through TaIrTe$_4$'s electronic structure. Angle-resolved photoemission spectroscopy measurements have revealed that the Fermi arcs extend along the k$_y$ direction with spin polarization primarily in the same direction.[29] When current flows along [100], this spin-momentum locked surface state naturally facilitates stronger charge-spin conversion, manifesting as enhanced NPHE and FL-SOT responses, similar to the mechanism observed in topological insulators.[10, 22, 24, 50]

However, the surface state contribution alone cannot fully explain the observed anisotropic behavior, particularly the NPHE signal along the [100] crystal direction, as the spin polarization on the quasi-one-dimensional Fermi arc has negligible k$_x$ component. Additional mechanisms must be considered, including possible bulk state contributions arising from Weyl points, TaIrTe$_4$'s naturally broken inversion symmetry and the presence of linear band crossing points.[30, 34] The strong anisotropy of DL-SOT, widely reported to originate from the bulk spin Hall effect,[57, 58] shows larger response along [010], contrasting with the FL-SOT behavior. This distinction is consistent with their different origins: while DL-SOT primarily arises from bulk spin-orbit coupling, FL-SOT is mainly attributed to interface spin accumulation, with its anisotropy correlating with the NPHE, suggesting the influence of surface spin-momentum locking.

**Conclusion**

In summary, we have successfully demonstrated the simultaneous observation of linear and nonlinear charge–spin conversions in a single TaIrTe$_4$/Py heterostructure system. By developing magnetic field-dependent harmonic Hall measurements, we achieved clear separation of NPHE, field-like SOT, and damping-like SOT contributions, enabling a comprehensive characterization of both linear and nonlinear charge-spin conversion properties. Our experiments revealed pronounced crystallographic anisotropy in TaIrTe$_4$, with distinct charge-spin conversion characteristics along [100] and [010] directions, reflecting the role of Fermi arc surface states and spin-momentum locking. The ability to simultaneously probe and distinguish between linear and nonlinear charge-spin conversion processes in TaIrTe$_4$ provides valuable insights into the fundamental mechanisms of spin current generation in Weyl semimetals. These findings not only deepen our understanding of anisotropic charge-spin conversion in topological materials but also suggest new strategies for optimizing spintronic device performance through crystal direction engineering.

**METHODS**

**Sample preparation and device fabrication**

The TaIrTe$_4$ single crystal was mechanically exfoliated and transferred onto a thermally oxidized silicon substrate with 500nm of SiO$_2$ by thermal release tapes. Then magnetron sputtering is used to deposit 7nm of Py onto the TaIrTe$_4$. To protect the heterostructure, we deposit 3nm of copper cap in situ onto the TaIrTe$_4$/Py bilayer. After deposition of the Py and copper layers, TaIrTe$_4$ flakes with high-quality surfaces were identified by optical microscopy. Selected flakes were then evaluated for thickness and surface homogeneity by atomic force microscopy. Finally, the sample was fabricated into a cross pattern with channel width of 8μm using a standard photolithography and Ar ion etching techniques.

**Second harmonic Hall measurements**

The second harmonic Hall signal was characterized using an in-plane harmonic lock-in technique. The measurement system uses Keithley 6221 as the AC current source and Stanford SR830 as the lock-in amplifier to measure electrical signals. The system is also equipped with a magnetic field and temperature control system, as well as a system for controlling the angle between the sample and the magnetic field. The $R_{xy}^{1\omega}$ and $R_{xy}^{2\omega}$ for an AC current I$_{ac}$ of 1333 Hz were simultaneously measured while rotating the sample in the plane (azimuthal angle φ) under an external field H$_{ext}$. All measurements were conducted at a temperature of 300K.

**ASSOCIATED CONTENT**

**Supporting Information**

NPHE and SOT results for charge current flows along TaIrTe$_4$ [010] direction, including Figures S1−S2.

## DATA AVAILABILITY

The data that support the findings of this study are available from the authors on reasonable request.

## COMPETING INTERESTS

The authors declare that there are no competing interests.

## ACKNOWLEDGEMENTS

This work was supported by the National Key Research and Development Program of China (Nos. 2019YFA0307800, 2017YFA0303600, 2021YFA1400401), the National Natural Science Foundation of China (Nos. 12174406, 11874367, 51931011, 52127803, 12274369, U22A6005, U2032204), the Key Research Program of Frontier Sciences, Chinese Academy of Sciences (No. ZDBS-LY-SLH008), K.C.Wong Education Foundation (GJTD-2020-11), the Ningbo Key Scientific and Technological Project (Grant No. 2022Z094).

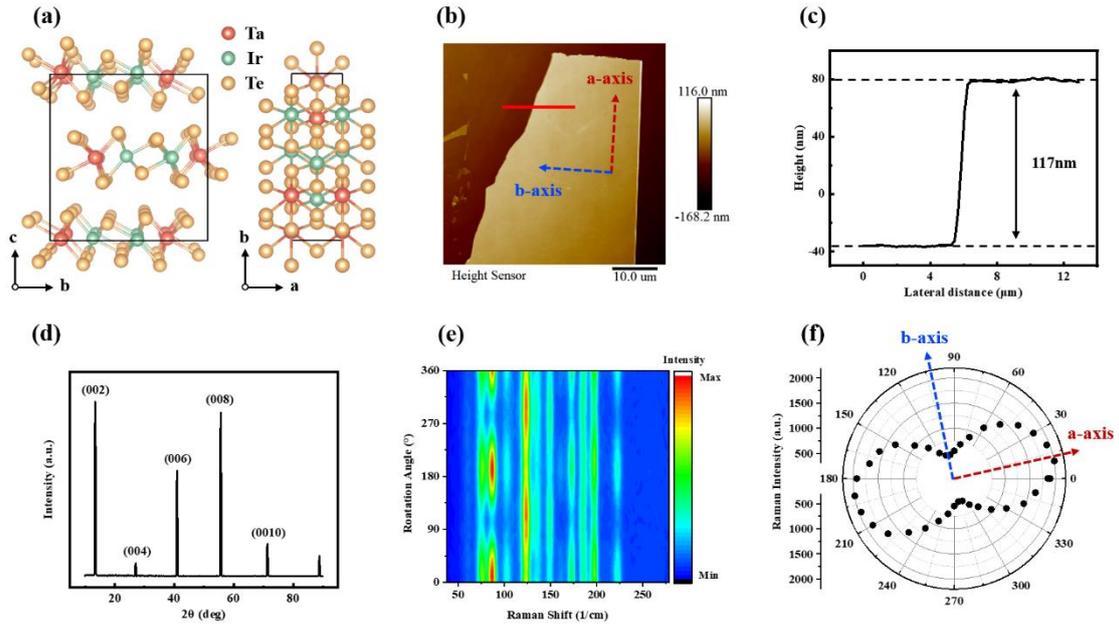

**Figure 1. Characterization of TaIrTe$_4$ single crystals.** (a) Crystal structure of TaIrTe$_4$ shown from the front (left) and top (right) views. (b) The AFM image of a mechanically exfoliated TaIrTe$_4$ flake transferred onto a SiO$_2$/Si substrate. (c) The height of the TaIrTe$_4$ flake along the red solid lines in (b). (d) X-ray diffraction pattern of the TaIrTe$_4$ single crystal. (e) The angle-dependent Raman intensity spectra of the TaIrTe$_4$ flake measured by rotating the sample in the parallel-polarized configuration. (f) Angular dependence of the intensity of the Raman spectra for the TaIrTe$_4$ flake at 86 cm$^{-1}$.

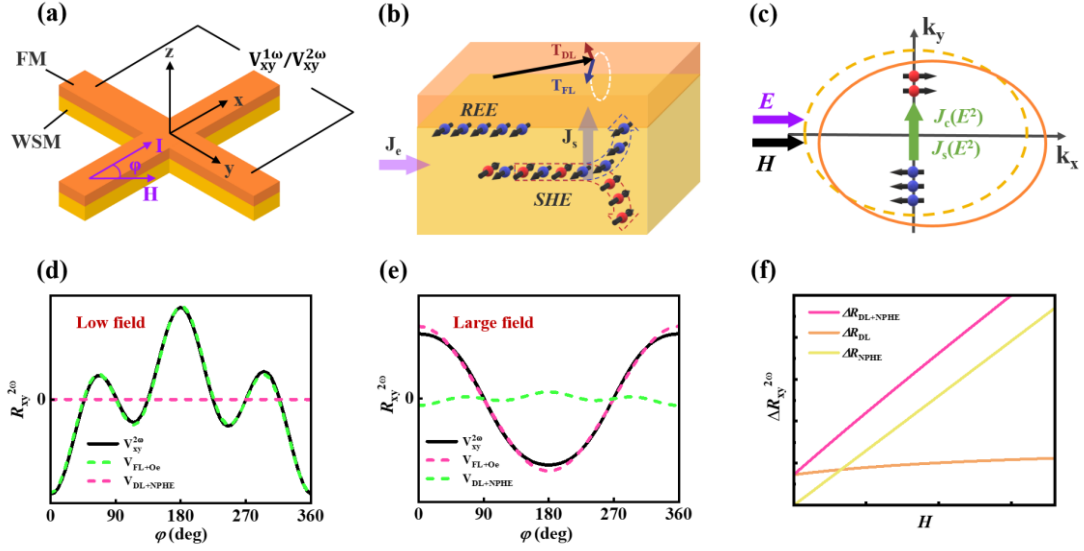

**Figure 2. Harmonic Hall measurements in the WSM/FM heterostructure.** (a) Schematic illustration of the device structure. φ denotes the angle between the current I and the in-plane magnetic field H. (b) Conceptual depiction of spin-orbit torque origins: bulk spin Hall effect (SHE) in the Weyl semimetal (WSM) and the interfacial Rashba–Edelstein effect (REE) at the WSM/FM interface. (c) Illustration of nonlinear change-spin conversion in the WSM layer, highlighting the generation of a transverse charge current $J_c(E^2)$ when the magnetic field H is applied parallel to the electric field E. This effect is detectable via nonlinear planar Hall measurements. (d, e) Angular dependence of the second harmonic Hall voltage of the WSM/FM bilayer at (d) low magnetic field and at (e) high field, with fitted curves showing contributions from the SOT and NPHE components. Note the dominance of different components at low and high magnetic fields. (f) The field dependence of the second harmonic Hall resistance components $\Delta R_{DL}$ and $\Delta R_{NPHE}$ on H, where both components follow a cosφ dependence.

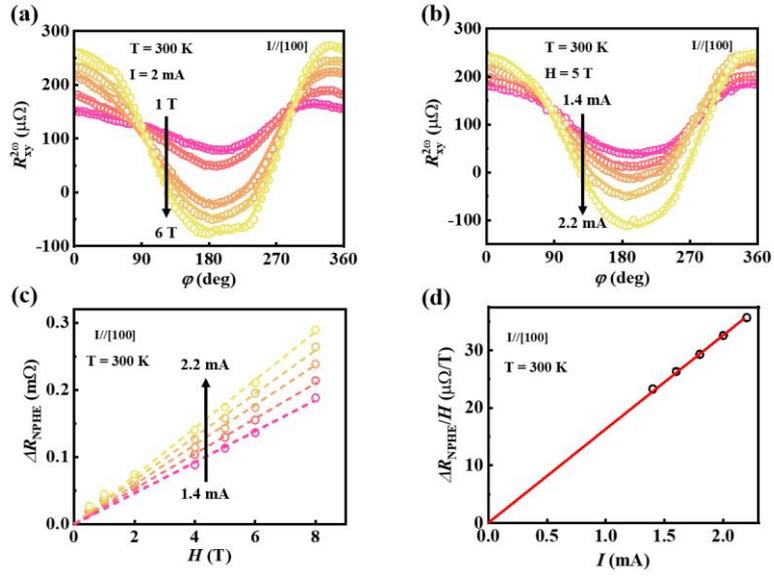

**Figure 3. Nonlinear planar Hall components from harmonic Hall measurements in the TaIrTe$_4$/Py bilayer (The current flows along the [100] direction).** (a), (b) Second harmonic Hall voltage as a function of φ under a fixed I of 2 mA with various H from 1 T to 6 T, and under a fixed H of 5 T with various I from 1.4 mA to 2.2 mA. (c) The dependence of the extracted nonlinear planar Hall resistance ΔR$_{NPHE}$ on H with various I from 1.4mA to 2.2mA. (d) The extracted ΔR$_{NPHE}$/H as a function of TaIrTe$_4$/Py heterostructure current I.

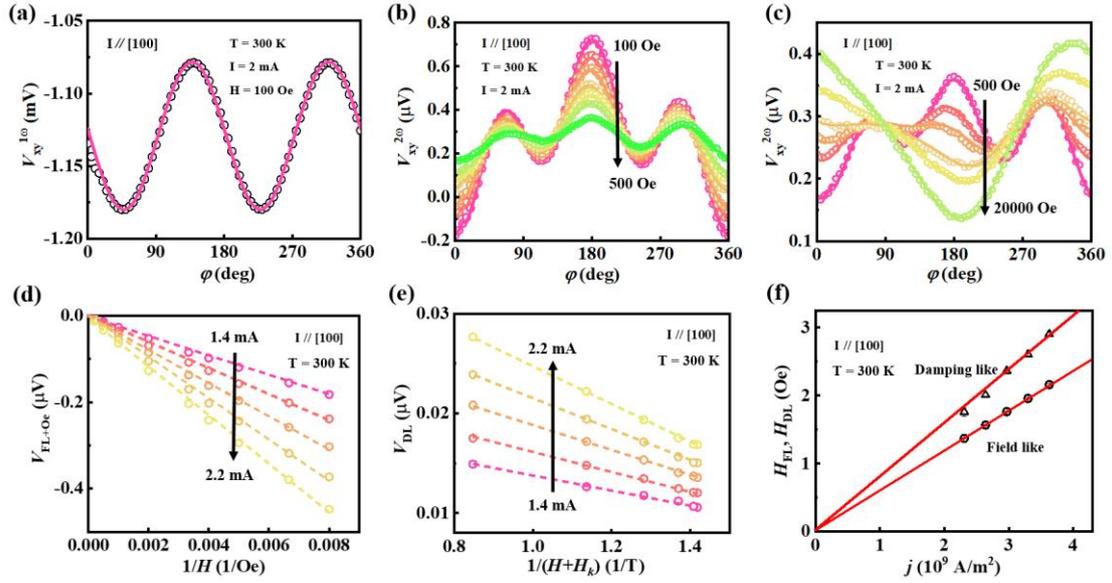

**Figure 4. Spin Orbit Torques components from Harmonic Hall measurements in the TaIrTe4/Py heterostructure. (The current flows along the [100] direction).** (a) The typical first harmonic Hall voltage of TaIrTe4/Py as a function of φ at I=2 mA and H = 100 Oe. (b), (c) The φ dependent second harmonic Hall voltage at I = 2 mA with various H from 100 Oe to 500 Oe (b) and from 500 Oe to 20000 Oe (c), the second harmonic hall voltage changes from cosφcos2φ-dependent to cosφ-dependent as the magnetic field increases. (d) The dependence of the extracted fieldlike voltage $V_{FL}$ for I from 1.4 to 2.2 mA on 1/H. (e) The extracted dampinglike voltage $V_{DL}$ plotted against $1/(H+H_k)$ for I from 1.4 to 2.2 mA, where $H_k$ is the anisotropy field of the Py layer. (f) The effective dampinglike and fieldlike fields, $H_{DL}$ and $H_{FL}$, as a function of the TaIrTe4 layer current density J.

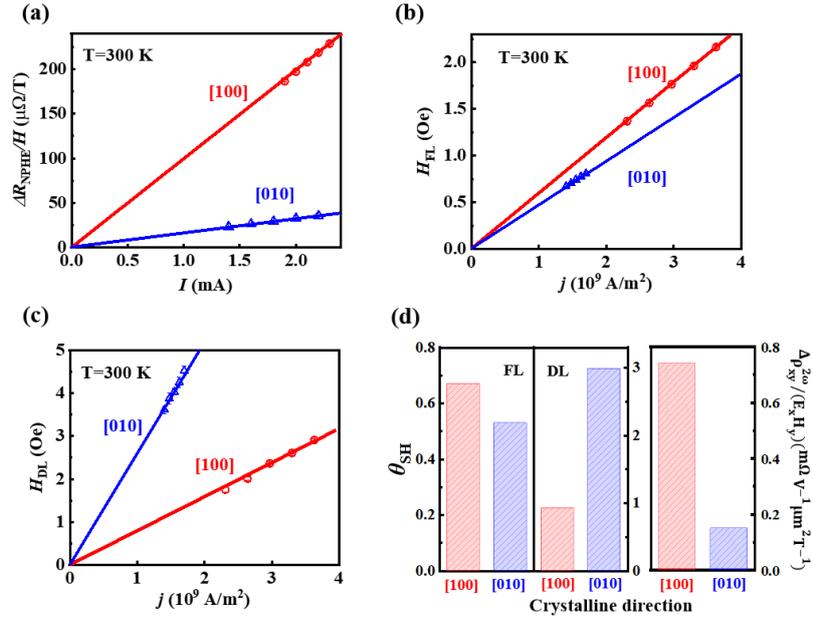

**Figure 5. Anisotropic NPHE and SOTs in TaIrTe$_4$.** (a) The extracted $\Delta R_{NPHE}$ normalized by the magnetic field H as a function of the TaIrTe$_4$/Py heterostructure current I for current applied along the TaIrTe$_4$[100] and TaIrTe$_4$[010] directions. (b), (c) The SOT effective field $H_{FL}$, $H_{DL}$ as a function of the TaIrTe$_4$ layer current density J for the [100] and [010] directions, respectively. (d) Summary of the FL and DL spin Hall angles $\theta_{SH}$ and the nonlinear Hall resistivity per unit electric and magnetic field $\chi$ for the [100] and [010] directions, revealing the strong anisotropy in both the linear and nonlinear charge-spin conversion efficiencies in TaIrTe$_4$.